# Perfecting the growth and transfer of large single-crystal CVD graphene: a platform material for optoelectronic applications.

V. Miseikis[1], S. Xiang[2], S. Roddaro[2], S. Heun[1], C. Coletti[1]

[1]*Center for Nanotechnology Innovation @NEST, Istituto Italiano di Tecnologia, Piazza San Silvestro 12, 56127 Pisa, Italy*

[2]*NEST, Istituto Nanoscienze—CNR and Scuola Normale Superiore, Piazza San Silvestro 12, 56127 Pisa, Italy*

*In this work we demonstrate the synthesis of millimetre-sized single-crystals of graphene, achievable in a commercially-available cold-wall CVD reactor, and several different approaches to transfer it from the growth substrate to a target substrate of choice. We confirm the high crystal quality of this material using various characterisation techniques, including optical and scanning electron microscopy as well as Raman spectroscopy. By performing field effect and quantum Hall effect measurements we demonstrate that the electronic properties of such single-crystals are comparable to those of ideal mechanically exfoliated flakes of graphene. Several applications of this high-quality material are also reviewed.*

## Introduction

Graphene grown by chemical vapour deposition (CVD) has emerged as the preferable type of material for large-scale applications [1]. The low cost of growth substrates and the high electronic and structural quality of CVD graphene have ensured its widespread adoption in research. It has been shown, however, that a major limiting factor of the quality of CVD graphene are grain boundaries [2], [3]. Therefore, in recent years a major research focussing point has been to increase the crystal size of CVD graphene [4], [5], [6], [7]. The key parameter for the growth of large-crystal graphene is the initial nucleation of graphene crystals on the highly-reactive copper substrates. Various approaches have been proposed in order to lower the nucleation density, including performing the growth on highly isolated (enclosed) surfaces [4], strongly diluting the methane precursor [6], or oxidising the copper surface to passivate it [7]. It should be mentioned that the approaches listed above typically require either very long growth times [6], or extensive customisation of the CVD systems [7]. Recently, fast growth of inch-sized single crystals of graphene was reported on Cu/Ni alloys [8], although a sophisticated CVD system with local precursor supply was used in this case.

In this work we present the development of a fast process for the synthesis of large-crystal graphene using a common, commercially available CVD system. We confirm the high structural and electronic quality of the transferred material by a variety of microscopy and spectroscopy techniques. Furthermore, we discuss the transfer of this material via different approaches: "wet" and "semi-dry" transfer. In fact, along with the charge carrier scattering by grain boundaries, transfer-induced contamination remains a limiting factor for the quality of CVD-grown graphene. The most common approach used for the transfer of graphene from copper and other catalysts involves chemical etching of the growth substrate using iron-based salts such as iron(III) chloride ($FeCl_3$) and iron(III) nitrate nonahydrate ($Fe(NO_3)_3$) [1], [9], or alternative "cleaner" etchants such as ammonium persulfate (($NH_4)_2S_2O_8$ also known as APS) [5], [10]. More recently, the so-called "bubbling" transfer technique has been proposed, whereby the graphene is detached from the growth substrate via electrochemical delamination without the need to etch the metal [11], [12]. We report electric measurements displaying the remarkable quality of large single-grain graphene transferred using the electrochemical delamination in combination with a dry-handling technique. We finally discuss and present selected optoelectronic applications of CVD graphene.

## Experimental

### CVD synthesis of graphene

Large-crystal graphene was synthesised on commercially-available copper foil with a thickness of 25 μm (99.8% purity, Alfa-Aesar, 13382), which prior to growth was electropolished as described previously [13] in order to remove the surface contamination and improve the surface flatness. This foil is also referred to later in the text as low purity foil (i.e., with a high oxygen content). Several growth experiments were also performed with higher purity foil (99.98%, Sigma-Aldrich) with equivalent surface treatment.

The synthesis was performed using an *Aixtron BM Pro* cold-wall reactor equipped with a 10 cm heater and dual-heating set-up (top and bottom). All growth experiments presented here were performed at a pressure of 25 mbar. A typical growth procedure consisted of 4 distinct parts: temperature ramp-up, annealing, growth, and cool-down (figure 1). A typical growth temperature was ~1060 °C. Temperature ramp-up (100 °C per minute) and annealing (duration: 10 min) was performed either in pure hydrogen or pure argon atmosphere, as will be discussed below. The growth (duration: 10-180 minutes, depending on the desired crystal size) was performed under a mixture of gas species (typical values of gas flow: 980 sccm Ar, 20 sccm $H_2$ and 1 sccm $CH_4$). The cool-down was per-formed in Ar/H₂ atmosphere (980 sccm / 20 sccm). To avoid accelerated oxidation of copper, the samples were unloaded from the CVD reactor at a temperature below 120 °C.

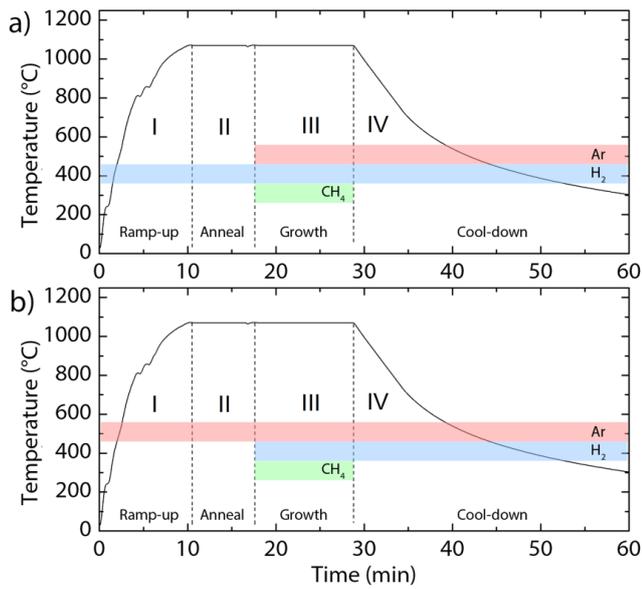

Figure 1. Process diagram indicating the temperature profile and gas flow for the process using (a) hydrogen annealing and (b) argon annealing.

Transfer

Two approaches were used to transfer graphene to the chosen substrates: "wet" and "semi-dry" transfer. "Wet" transfer was performed by first spin-coating the Cu/graphene sample with a PMMA support layer. Then the copper substrate was etched using an appropriate etchant (0.1M $FeCl_3$ or 0.44 M (100 g/L) APS, Sigma-Aldrich), leaving the graphene/PMMA membrane floating on the surface of the etchant solution. The membrane was transferred to several baths of deionised water in order to rinse it thoroughly. The target substrate was then used to pick up the membrane from the deionised water, and the stack was dried in ambient conditions. Finally, the PMMA film was removed in acetone and isopropanol (IPA).

For the "semi-dry" transfer, the samples were initially spin coated with a PMMA support layer. A semi-rigid support frame made using polyimide adhesive tape (3M) was attached around the perimeter of the foil. To minimise contamination, chemical etching of the Cu substrate was not performed; instead, the graphene/PMMA stack was removed from the growth substrate by electrochemical delamination, using sodium hydroxide (NaOH) as electrolyte. The graphene/PMMA membrane (suspended and stretched flat by the support frame) was rinsed in deionised water, dried, and deposited on the target substrate. To improve the adhesion, the samples were heated 10 minutes at 120 °C. Finally, the PMMA was removed in acetone and IPA.

Characterisation

The samples were analysed using a combination of microscopy and spectroscopy techniques. SEM imaging was performed using a *Zeiss Merlin* column with an accelerating voltage of 5 kV. Raman spectroscopy was performed using a *Renishaw inVia* system equipped with a 532 nm laser and a motorised stage for 2D mapping of samples.

The transport properties of large single-crystal graphene samples were investigated by conducting field effect and quantum Hall effect measurements. Isolated crystals were transferred to pre-patterned substrates of highly n-doped silicon with a 300 nm layer of $SiO_2$, which was chosen to allow back-gating of graphene devices. Subsequently, electron-beam lithography was used to contact the individual crystals in a 6-terminal configuration. The measurements were performed using a lock-in technique, utilising a Heliox Helium-3 cryostat with base temperature of 250 mK.

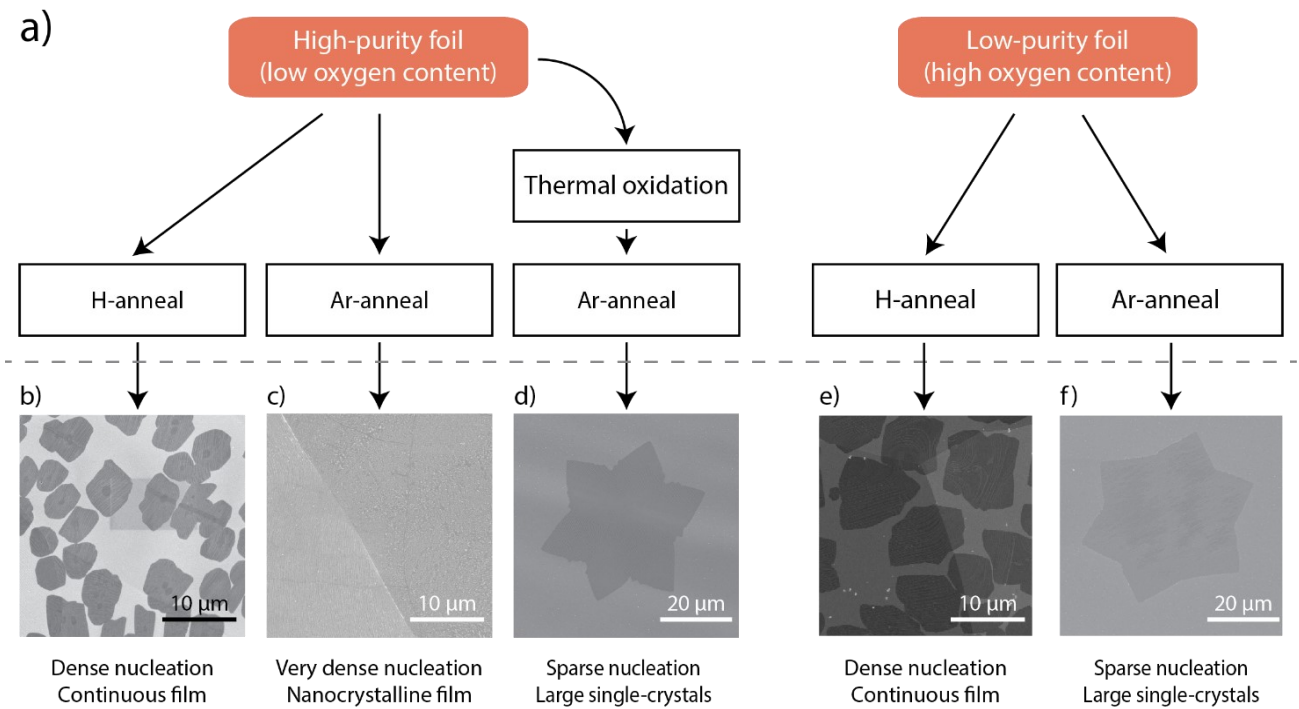

**Figure 2.** a) Process diagram of growth experiments performed with two kinds of copper foil. (b-f) SEM images of partially grown samples employing the different substrates and growth conditions. Note the different scale bars used for large-crystal images.

# Results and discussion

## Hydrogen annealing vs argon annealing

During the initial growth experiments, the ramp-up and annealing steps were performed in a pure hydrogen atmosphere. At high temperatures hydrogen is a strong reducing agent, helping to clean the Cu substrates from oxides and other impurities. This is a commonly-used approach for the growth of polycrystalline films, however, due to a high nucleation density, it is unsuitable for synthesising large crystals of graphene. Growth experiments performed on copper foil with both low and high oxygen content yielded similar results (figure 2 (b) and (e), respectively) confirming nucleation densities as high as 10 000 per $cm^2$.

Recently it has been demonstrated that the presence of oxygen in the Cu substrate has a strong effect on the nucleation density and growth dynamics of CVD-graphene, making oxidised substrates desirable for large crystal-growth [7]. Due to safety concerns of introducing oxygen into a system containing highly combustible gases, as was done in the work cited above, we instead utilised the native oxides present in the low-purity Cu substrates [13]. To prevent the reduction of these oxides, annealing was performed in an inert argon atmosphere instead of hydrogen. Growth experiments performed using low-purity foil annealed in argon revealed a decreased nucleation density on the order of ~1000 grains per $mm^2$, an improvement of one order of magnitude. Additionally, annealing in argon provided an increased copper grain size of several millimetres.

In order to confirm that the presence of oxygen is the main factor limiting the nucleation density, Ar-annealing experiments were also performed using substrates with low oxygen content. As expected, this did not yield low nucleation density - on the contrary, the surface of such samples was often observed to be highly reactive, causing dense and uncontrollable nucleation and subsequent formation of nanocrystalline films (figure 2 (c)). However, by simply heating the low-oxygen foil for 2 minutes at 180 °C in ambient atmosphere, the surface of such substrates could be intentionally oxidised. Subsequent growth experiments on such thermally oxidised foil provided qualitatively similar results to those obtained on natively oxygen-rich foil (figure 2 (d)).

To further reduce the gas impingement flux on the sample, the foil (either low purity or thermally oxidised high purity) was contained within a small-volume custom-made enclosure, consisting of a quartz disk placed on 6 mm-thick graphite spacers. Such sample containment provided a significant further reduction of nucleation density to approximately 15 grains per $mm^2$. With such nucleation density, the growth time could be extended up to 1 hour while retaining the isolated large crystals with size of up to 750 µm. In general, statistical analysis of many growth experiments revealed a growth rate of ~15 µm/min (figure 3 (b)), among the highest reported to date. It is noteworthy that the use of an enclosure was beneficial only in combination with non-reducing argon-annealing and did not have a strong effect of lowering the nucleation density on the highly reactive surface of Cu foil obtained using the hydrogen-annealing process.

To further decrease the nucleation density and allow synthesis of millimetre-sized single-crystals, the foils were folded into a pocket-like enclosure, initially proposed by Li *et al.* [4] and shown in Fig. 4 (a). Due to the crimped edges of such enclosures with a practically airtight seal, the inside Cu surfaces of these "pockets" have little exposure to the carbon precursor (mainly due to the diffusion of carbon species through the copper [14]). This leads to extremely low nucleation densities of well below 1 crystal per $mm^2$, allowing to extend the growth time up to 3 hours and to synthesise isolated single crystals of up to 3.5 mm (Fig. 4 (b)). In the following sections we discuss the transfer, properties, and applications of large single-grain graphene obtained in enclosed and oxidised low-purity Cu foil.

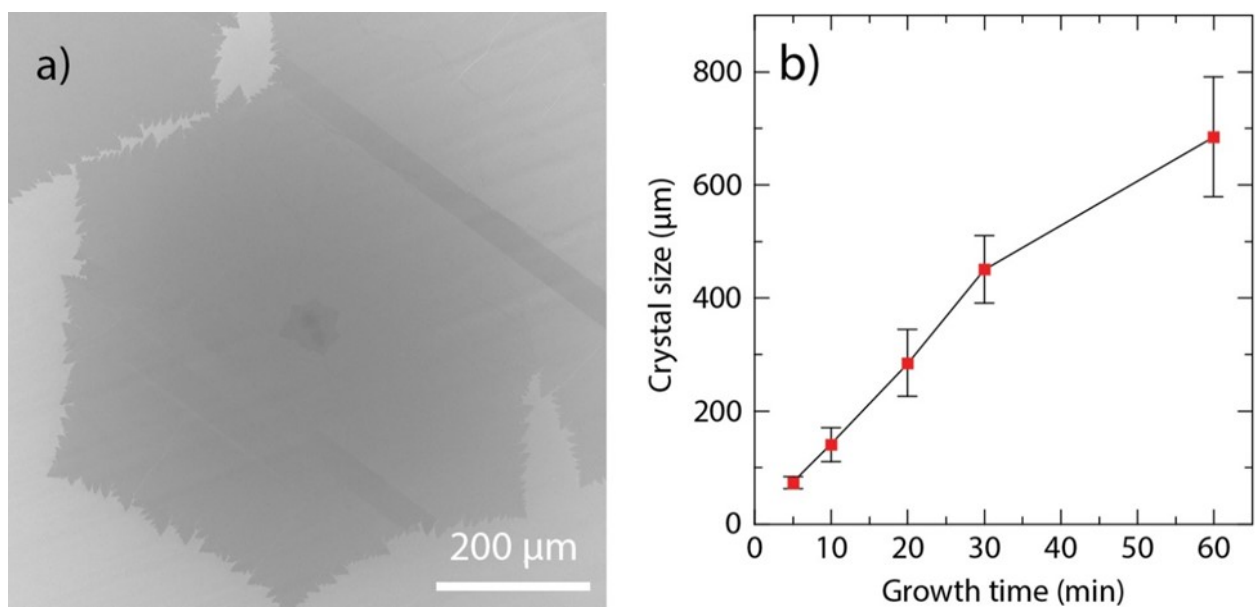

**Figure 3.** a) Single-crystal obtained after 1 hour of growth. b) Crystal size dependence on the growth time.

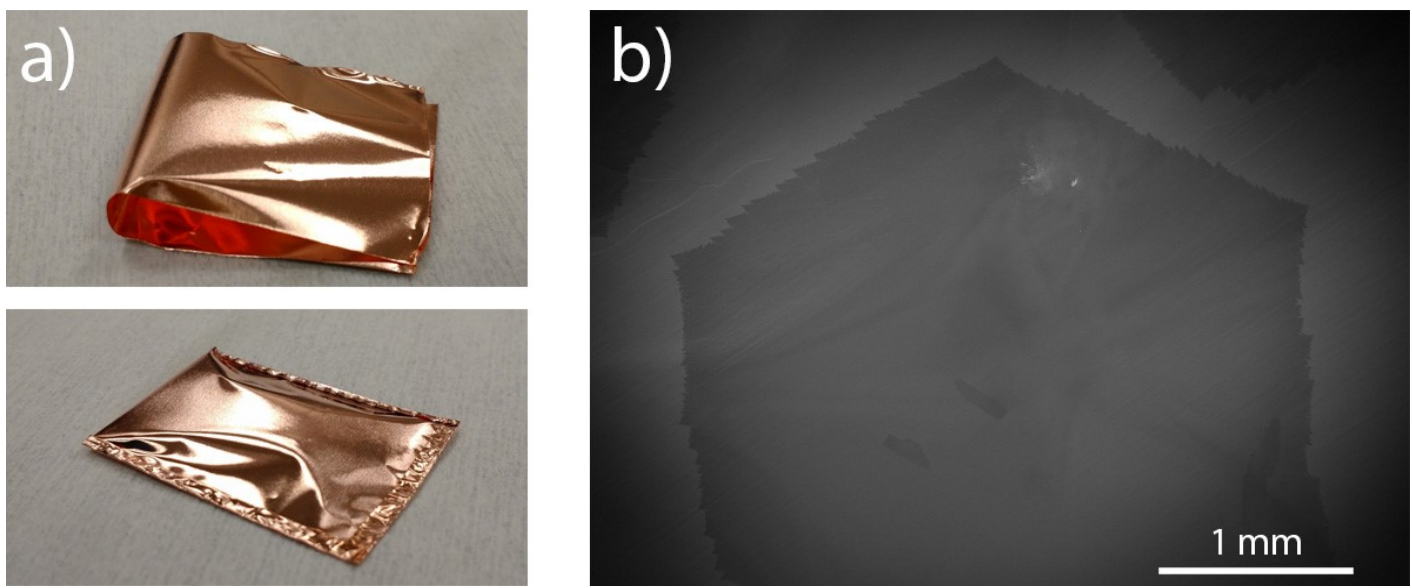

**Figure 4.** a) Formation of the copper "pocket": the foil is folded over (top), and the edges are crimped to create an airtight enclosure (bottom). b) Single-crystal of graphene with a diameter of more than 3 mm grown inside the "pocket" enclosure.

## Optimisation of graphene transfer for contamination-free graphene

As described in the experimental section, wet etching with iron(III) chloride and APS as well as electrochemical delamination were employed for the transfer of large grain graphene on various substrates. It should be noted that in the case of electrochemical delamination, we used a modified approach utilising a semi-rigid frame for dry handling of graphene/PMMA membranes. Notably, adoption of a frame allows attachment of the membrane to a micromanipulator stage for precise (on the order of a few µm) alignment of graphene single crystals on the target substrate.

Aside from microscopy techniques, a quantitative measure of the quality and cleanliness of transferred graphene is the analysis of the 2D peak in the Raman spectrum of graphene [15], [16]. In particular, the width of the 2D peak $\Gamma_{(2D)}$ is an indicator of strain variations within the laser probe spot, which can also be responsible for charge carrier scattering, and hence is a major limitation of electronic mobility. Comparing the Raman spectra obtained from our samples prepared using the different transfer techniques, clear differences in the width of the 2D peak were consistently observed, even though the defect-related D peak was rarely observed in any of the samples, indicating the overall high crystal quality of graphene. $\Gamma_{(2D)}$ after $FeCl_3$ transfer was typically found to be ~35 cm$^{-1}$. When using APS, the typical $\Gamma_{(2D)}$ values were ~29 cm$^{-1}$. For the samples prepared using "semi-dry" transfer of electrochemically-delaminated graphene, a typical $\Gamma_{(2D)}$ value was 27 cm$^{-1}$, comparable to that of mechanically exfoliated flakes. Typical Raman spectra obtained in each case are shown in figure 5.

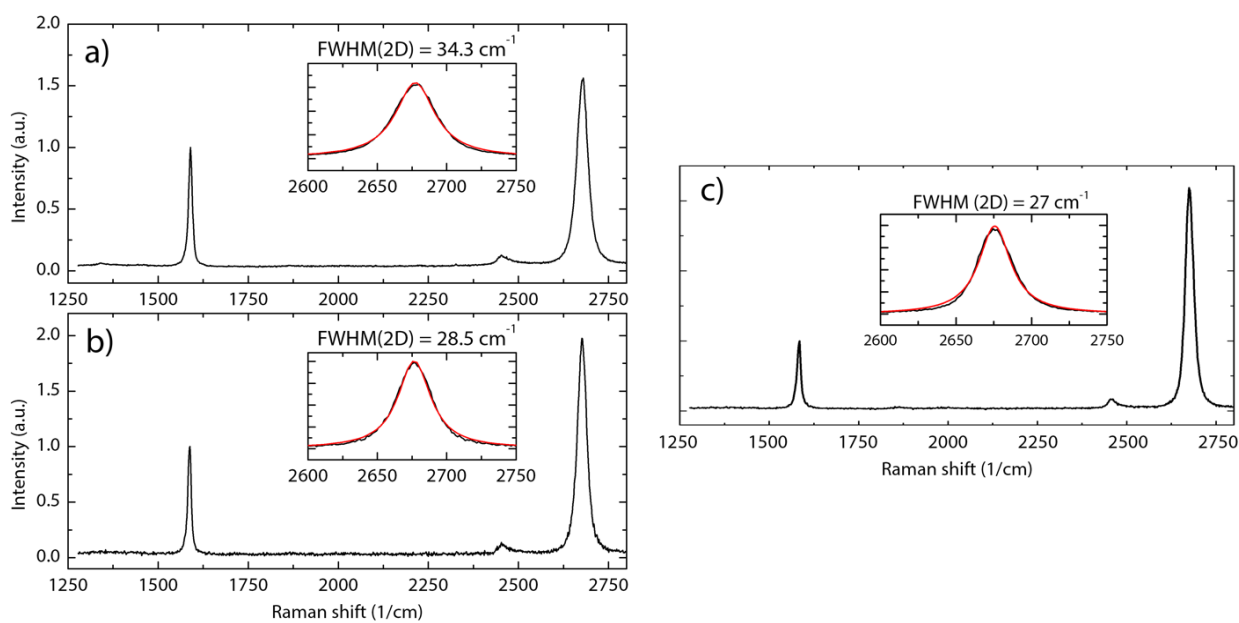

**Figure 5.** Typical Raman spectra obtained from samples prepared using different transfer approaches. a) Cu etching by $FeCl_3$, b) Cu etching by APS, c) electrochemical delamination and semi-dry transfer. Inset of each graph shows a Lorentz fitting of the 2D peak.

The "semi-dry" transfer undoubtedly provided the lowest impurity levels, thanks to the self-cleaning effect of the hydrophobic membranes. However, some samples transferred with this approach suffered adhesion issues. In particular, extra-large grains with a diameter above 500 μm and continuous films of graphene typically contained numerous pinholes. The formation of such defects could be explained by the trapping of small air pockets between the substrate and the graphene membrane, which is known to be impermeable to gas [17]. When the samples with freshly-transferred graphene are placed in acetone to remove the PMMA support, the trapped air can escape by breaking the graphene and causing a pinhole. For smaller grains, the statistical probability of trapping air was much lower, and far fewer damaged crystals were observed. On the other hand, during "wet" transfer, a thin and bubble-free layer of water was typically present at the graphene/substrate interface, which was subsequently slowly removed by the capillary effect, ensuring homogeneous adhesion of graphene. For this reason, when large graphene grains (with a diameter above 300 μm) need to be transferred, the wet transfer approach remains the preferable option, despite the relatively higher amount of contamination. Further work is needed to optimise the "semi-dry" transfer method and improve the adhesion of graphene on target substrates over large areas.

## Measurement of transport properties

Transport measurements were conducted in graphene single-crystals transferred with the "semi-dry" approach and contacted in a 6-terminal configuration, as shown in figure 6. A source-drain current of 10 nA was applied across the graphene crystal, and the side contacts were used to measure the longitudinal voltage drop $V_{xx}$ and the transverse (Hall) voltage drop $V_{xy}$.

Electric field effect was measured by sweeping the back-gate voltage in the range of -45 to 60 V. A clear and sharp resistivity peak was observed at a back gate voltage of around +8.5 V, indicating an impurity-induced charge density of approximately 6 x $10^{11}$ $cm^{-2}$, a relatively low value for graphene on Si/$SiO_2$. Furthermore, fitting the field effect measurement data using the Drude model (figure 7 (a)) suggests carrier mobility above 10 000 $cm^2$/Vs, which is comparable to samples prepared on $SiO_2$ using pristine, mechanically exfoliated flakes.

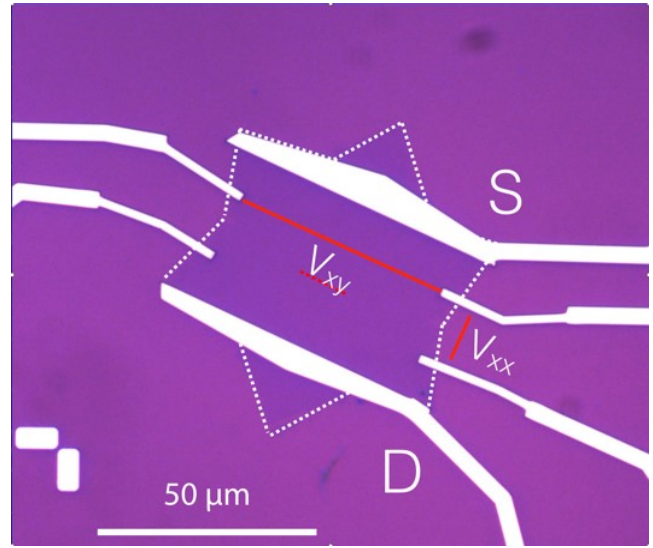

**Figure 6.** Optical image of a large crystal of graphene contacted for transport measurements.

Quantum Hall measurements were performed by applying a magnetic field of up to 10 T. Remarkably, even though a non-ideal Si/$SiO_2$ substrate was used, upon application of magnetic field, pronounced oscillations of $V_{xx}$ and plateaus of $V_{xy}$ were observed, clearly indicating the quantum Hall effect. At a field of 4 T, up to 10 well-defined levels could be distinguished (figure 7(b)), further confirming the exceptionally high quality of the samples. A more thorough representation of the quantum Hall measurement of these samples is shown in the form of a Landau fan diagram (Fig. 8).

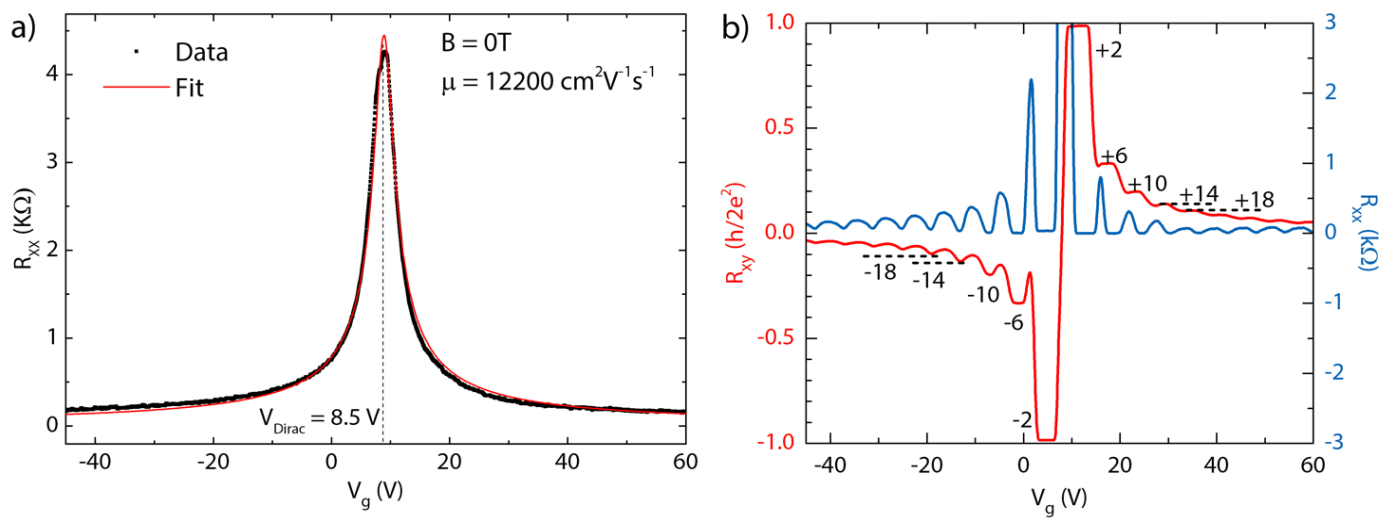

**Figure 7.** a) Electric field effect at zero applied magnetic field. b) Traces of longitudinal resistance $R_{xx}$ (blue) and transverse (Hall) resistance $R_{xy}$ (red) as a function of $V_g$ obtained at B=4T.

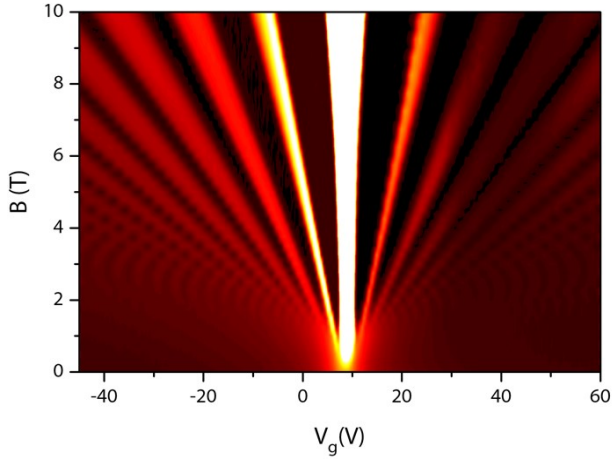

**Figure 8.** Landau fan diagram of magnetoresistance $R_{xx}$ as a function of $V_g$ and B.

Furthermore, quantum interference was studied in these samples [18], by analysing the weak localisation peak at various back-gate voltages and temperatures, revealing a carrier dephasing length comparable to that of exfoliated graphene and larger than that previously observed in CVD-grown graphene.

The field effect and quantum Hall effect measurements presented above clearly demonstrate that the electric quality of single-crystal CVD graphene is comparable to that of mechanically exfoliated flakes. It should be noted that the measured mobility values were likely to be limited by the non-ideal Si/$SiO_2$ substrate. Previous work suggests that further improvements in transport characteristics of such samples of CVD graphene can be achieved by using high-quality substrates such as hexagonal boron nitride [5], [16].

## Applications

The high quality of large single-crystal CVD graphene has made it attractive for various applications. Recently, Spirito et al used this material to fabricate UV-sensitive photodetectors [19]. The devices contained graphene field effect transistors coated with a sensitizing layer of colloidal CdS nanocrystals – i.e. the nanocrystals absorbed the incident light creating electron-hole pairs; and the electrons were then transferred to graphene, generating a detectable photocurrent. The detectors demonstrated high responsivity of up to $10^4$ A/W and a fast response rate in the kHz range.

The high quality of graphene was even more important in a recent investigation of transport in strongly-coupled graphene and $LaAlO_3$/$SrTiO_3$ oxide (LAO-STO) junctions [20]. In order to maintain a pristine, contamination-free interface between the graphene single-crystals and LAO-STO, the "semi-dry" transfer technique was employed during the fabrication of these samples. At low applied bias between the graphene and LAO-STO, strong interlayer electrostatic coupling was observed, allowing the gating of graphene. On the other hand, at higher vertical bias above |1V|, direct tunnelling coupling was measured.

As discussed above, large crystals of CVD-graphene can demonstrate electric properties comparable to those measured using mechanically exfoliated flakes. However, in certain cases where the high mobility of graphene is not a necessary prerequisite, polycrystalline films prepared using the hydrogen annealing could provide significantly larger continuous areas of graphene than are possible to achieve with single-crystals. One application of such material was a hybrid metasurface consisting of terahertz split-ring resonator (SRR) arrays and graphene, investigated by Zanotto et al [21]. Wet transfer was used to deposit a sheet of graphene on GaAs substrates containing a gold split-ring resonator metamaterial and a 30 nm insulating layer of $SiO_2$. The presence of graphene modified the THz transmission response of the metamaterial, allowing its modulation by an applied magnetic field. Notably, a magnetic field of 5 T applied to a hybrid graphene/SRR metasurface caused transmittance modulation of up to 10%, compared to just ~1% modulation of unpatterned graphene/$SiO_2$/GaAs stack, indicating a significant interaction between the Dirac fermions in graphene and the SRR resonance. Furthermore, numerical simulations of magneto-optic response of such hybrid material were performed, with good agreement to the experimental data.

## Conclusions

We have presented a route to obtain high-quality single-crystal graphene using a commercially available CVD system. Using a variety of characterisation techniques, we confirm the high quality of the material, and we present several approaches to integrate this material on different substrates of choice. We demonstrate that using state-of-the-art transfer techniques, it is possible to fabricate large-area devices with electric quality comparable to that of mechanically exfoliated graphene. Finally, we present some recent applications of our material.

## Acknowledgements

The authors acknowledge financial support from the Italian Ministry of Foreign Affairs (Ministero degli Affari Esteri, Direzione Generale per la Promozione del Sistema Paese) in the framework of the agreement on scientific collaborations with Canada (Quebec) and Poland; and from the CNR in the framework of the agreement on scientific collaborations between CNR and JSPS (Japan), CNRS (France), and RFBR (Russia). We also acknowledge funding from the European Union Seventh Framework Programme under grant agreement no. 604391 Graphene Flagship.